\documentclass[doublecol]{epl2} 

\usepackage{amsmath,amssymb,braket,bm,dsfont}
\usepackage[mathscr]{euscript}
\usepackage{graphicx}       

\newcommand{\Co}{C_\text{o}}
\newcommand{\Cnn}{C_\text{nn}}
\newcommand{\tr}{\text{tr}\,}

\title{Restoring number conservation in quadratic bosonic Hamiltonians with dualities} 
\shorttitle{A duality for quadratic bosonic Hamiltonians} 

\author{Vincent P. Flynn\inst{1}\and Emilio Cobanera\inst{2,1},\and Lorenza Viola\inst{1}}
\shortauthor{Vincent P. Flynn \etal}

\institute{                    
\inst{1}Department of Physics and Astronomy, Dartmouth
College - 6127 Wilder Laboratory, Hanover, NH 03755, USA \\
\inst{2}Department of Mathematics and Physics, SUNY Polytechnic Institute - 100 Seymour Ave., Utica, NY 13502, USA}
  

\abstract{Number-non-conserving terms in quadratic bosonic Hamiltonians can induce unwanted dynamical instabilities. By exploiting the pseudo-Hermitian structure built in to these Hamiltonians, we show that as long as dynamical stability holds, one may always construct a non-trivial dual (unitarily equivalent) number-conserving quadratic bosonic Hamiltonian. We exemplify this construction for a gapped harmonic chain and a bosonic analogue to Kitaev's Majorana chain. Our duality may be used to identify local number-conserving models that approximate stable bosonic Hamiltonians without the need for parametric amplification and to implement non-Hermitian $\mathcal{PT}$-symmetric dynamics in non-dissipative number-conserving bosonic systems. Implications for computing topological invariants are addressed. }

\begin{document}
\maketitle

\section{Introduction}
\label{intro}
Duality transformations provide a powerful conceptual framework for relating seemingly very different physical systems or very different regimes (thermodynamic phases) of a single system. One paradigmatic example is the duality that relates the properties of the Ising model on the square lattice at low and high temperatures \cite{Wannier}.  Generally, dualities  are transformations of the microscopic degrees of freedom which preserve, in a suitable sense, the locality structure and spectral properties of the original Hamiltonian \cite{savit,DPRL,BAD} -- though possibly nothing else. A duality can change drastically the symmetries of a system in that, for example, a global symmetry of the original system may be mapped to a boundary symmetry of its dual or, a broken number-conservation symmetry may be mapped to a broken translation symmetry \cite{GaussianD}. Of all the typical, but not quite defining, features of duality transformations (non-locality, mapping microscopic to topological degrees of freedom, or strongly-coupled systems to weakly-coupled ones), this phenomenon of \textit{symmetry transmutation} is arguably their most universal feature. 

In this paper, we show that a large class of number-non-conserving quadratic bosonic Hamiltonians (QBHs) are dual to number-conserving QBHs in a conceptually and practically useful sense: one can always recast and simulate these QBHs, which feature ``pairing'' or  ``parametric amplification'', in terms of unitarily equivalent QBHs, which \emph{feature only hopping terms}, by properly adjusting their amplitudes. In the process, the total number operator undergoes the transmutation typical of dualities. Our dualities are akin to the \emph{Gaussian dualities} connecting free-fermion topological insulators and superconductors \cite{GaussianD,PRB2}. The specification of the class of QBHs that admit number-conserving duals is in itself remarkable. Unlike free fermions, a QBH can be thermodynamically unstable by failing to be bounded below and/or it can be dynamically unstable by inducing non-periodic, unbounded time evolution of some observables. The class of number-non-conserving QBHs that are dual to number-conserving ones is precisely the class of \emph{dynamically stable QBHs}. A QBH can show transitions between dynamically stable and unstable regimes as a function of Hamiltonian parameters and,  as we showed in \cite{decon}, these transitions are characterized by the breaking of a generalized parity-time ($\mathcal{PT}$) symmetry, with sharp features emerging in the thermodynamic limit. Thus, the notion of ``dynamical phase diagram'' is granted for bosons and distinct from the usual notion of thermodynamic phase diagram. Our dualities identify number-conserving dual QBHs within the dynamical phase diagrams of number-non-conserving QBHs.

Beside introducing the general construction and elucidating basic features of our duality transformations, we provide two prominent examples, namely, the number-conserving duals of a gapped harmonic chain and the bosonic Kitaev-Majorana chain of under various boundary conditions \cite{clerkBKC,decon}. Furthermore, we highlight two important applications of our duality. First, we derive a formula that relates the geometric phase and other topological invariants of number-non-conserving QBHs to the corresponding (Berry) invariants of the dual number-conserving system.  Second, we outline an approach for realizing the $\mathcal{P}\mathcal{T}$-symmetric dynamics characteristic of many relevant semiclassical open systems with balanced gain and loss in terms of number-conserving QBHs. Thus, not only can the introduction of non-unitary noise effects be avoided, as advocated in \cite{clerkPT}, but the need for precisely engineering pairing/parametric amplification may be bypassed altogether. We believe this approach could offer significant advantages for analog quantum simulation \cite{nori}, and for furthering the exploration of the remarkable physical phenomena associated with non-Hermitian systems \cite{PT}. 

\section{Background: The effective BdG Hamiltonian} 
\label{background}
The class of QBHs we consider are of the form
\begin{equation}
\label{genham}
\widehat{H}=\sum_{i,j=1}^N \left[K_{ij} a_i^\dag a_j + \frac{1}{2}\left(\Delta_{ij} a_i^\dag a_j^\dag + \Delta_{ij}^*a_ja_i\right)\right], 
\end{equation}
where $a_i$ ($a_i^\dag$) is the bosonic annihilation (creation) operator for mode $i$, with $[a_i,a_j]=0$, $[a_i,a_j^\dag]=\delta_{ij} 1_F$, and $1_F$ being the identity on the bosonic Fock space. Hermiticity of $\widehat{H}$ implies that $K^\dag=K$ and bosonic commutation rules allow us to take $\Delta^T=\Delta$. By introducing a Nambu array $\hat{\Phi}\equiv [a_1,a_1^\dag,\ldots,a_N,a_N^\dag]^T$, we may formally define a Hermitian single-particle Hamiltonian (SPH) $H$ such that
\begin{equation}
\label{genham2}
\widehat{H} = \frac{1}{2}\hat{\Phi}^\dag H \hat{\Phi}-\frac{1}{2}\text{tr} K, \quad 
[H]_{ij}= \begin{bmatrix}
K_{ij} & \Delta_{ij} \\ \Delta^*_{ij} & K^*_{ij} .
\end{bmatrix}, 
\end{equation}
The QBH \(\widehat{H}\) is bounded below (thermodynamically stable) if and only if \(H\) is positive-semidefinite \cite{Derezinski17}. 

Let  $\ket{\alpha(t)}$ denote a vector in the Hilbert space $\mathcal{H}\equiv (\mathbb{C}^{2N},\braket{\cdot|\cdot})$ and $\tau_j\equiv\mathds{1}_N\otimes \sigma_j$ in terms of the usual Pauli matrices. The Heisenberg equations of motion for an operator of the form $\widehat{\alpha}(t)\equiv \bra{\alpha(t)}\tau_3 \hat{\Phi}(0)$ are
\begin{equation}
\label{Heis}
i\frac{d}{dt}\widehat{\alpha}(t) = - [\widehat{H},\widehat{\alpha}(t)] = \widehat{G\alpha}(t),\quad G\equiv\tau_3 H, 
\end{equation} 
which further reduces to the  linear time-invariant system $\ket{\dot{\alpha}(t)} = iG\ket{\alpha(t)}$. Hence, the {\textit{effective Bogoliubov-de Gennes (BdG) SPH} $G$ completely characterizes the dynamics of physical observables.
 
The matrix \(G\) is Hermitian if and only if the pairing amplitudes $\Delta_{ij}$, which break the conservation of the total number operator \(\widehat{N}\equiv\sum_{i=1}^Na_i^\dagger a_i\), vanish. Nonetheless, \(G\) has two built-in symmetries: (i) $\tau_3$\textit{-pseudo-Hermiticity}\footnote{Recall that a matrix $M$ is called $\eta$-pseudo-Hermitian if there exists a Hermitian, invertible matrix $\eta$ such that $M = \eta M^\dag \eta^{-1}$ \cite{ali}.} ($\tau_3$-PH), that is, $G^\dag = \tau_3 G\tau_3$, and (ii) \textit{charge conjugation symmetry}, that is, $G=-\mathcal{C}G\mathcal{C}^{-1}$, with $\mathcal{C}=\tau_1\mathcal{K}=\mathcal{C}^{-1}$ and $\mathcal{K}$ complex conjugation.  As a consequence, the eigenvalues of \(G\) come generically in quartets, \(\{\omega, -\omega^*, -\omega, \omega^*\}\). One can understand these features geometrically. On the one hand, because of $\tau_3$-PH, $G$ is `Hermitian' in the indefinite inner-product \emph{Krein space} ${\mathscr K}_{\tau_3} \equiv\left( \mathbb{C}^{2N},\braket{\cdot|\tau_3|\cdot}\right)$\cite{baldesflow}. On the other hand, given two operators $\widehat{\alpha}=\bra{\alpha}\tau_3 \hat{\Phi}$ and $\widehat{\beta}=\bra{\beta}\tau_3\hat{\Phi}$, we have that $[\widehat{\alpha},\widehat{\beta}^\dag ] =\braket{\alpha|\tau_3|\beta}$. Therefore, the commutation relations of the normal modes of \(\widehat{H}\) are also determined by the \(\tau_3\)-inner product along with the identity $\widehat{\alpha}^\dag = -\widehat{\mathcal{C}\alpha}$ in terms of the charge-conjugation operation \({\cal C}\).

The focus of this paper is on QBHs such that the effective SPH $G$ is {\em diagonalizable with a real spectrum}. These  conditions imply that the time evolution of the normal modes of Eq.\,\eqref{Heis} is bounded (quasi-periodic) and so the system is \textit{dynamically stable} -- see \cite{decon} for an in-depth analysis of the dynamical phases and the corresponding phase boundaries in general QBHs. With these assumptions, we are guaranteed the existence of an eigenbasis $\{\ket{\psi_n}\}$ of $G$, with corresponding eigenvalues $\omega_n\in\mathbb{R}$, that can be chosen to satisfy (i) $\ket{\psi_{n+N}} = \ket{\overline{\psi}_n} = \mathcal{C}\ket{\psi_n}$, and (ii) $\braket{\psi_n|\tau_3|\psi_m}=\delta_{nm}$ for $n\leq N$ and $-\delta_{nm}$ for $n>N$ \cite{ripka,RK}. The many-body Hamiltonian $\widehat{H}$ can thus be cast as a sum of independent simple harmonic oscillators 
\begin{equation}
\label{decoupled}
\widehat{H} = \sum_{n=1}^N \omega_n \Big(\widehat{\psi}_n^\dag \widehat{\psi}_n+\frac{1}{2}\Big)-\frac{1}{2}\text{tr}K , 
\end{equation}
where $\widehat{\psi}_n = \bra{\psi_n}\tau_3\hat{\Phi}$, $[\widehat{\psi}_n ,\widehat{\psi}_m^\dag] = \delta_{nm}$ and \mbox{$[\widehat{\psi}_n,\widehat{\psi}_m] = 0$}. The quasi-particle vacuum can be constructed in the standard manner \cite{ripka,RK} and coincides with a ground state of \(\widehat{H}\) provided that the \(\omega_n\geq 0\). We do not require that this should be the case.

\section{Restoring Hermiticity of the effective BdG Hamiltonian}
\label{Smatrix}
Drawing from general results on PH operators \cite{ali}, it follows that dynamical stability guarantees the existence of a \emph{positive-definite matrix $S$} with the property $G=S^{-1}G^\dag S$. Thus, a dynamically stable effective BdG Hamiltonian \emph{is} Hermitian when regarded as an operator on the Hilbert space ${\cal H}_S\equiv (\mathbb{C}^{2N},\braket{\cdot|S|\cdot})$. This Hermitian inner product can be characterized in terms of the eigenbasis $\{\ket{\psi_n}\}_{n=1}^{2N}$ described right above Eq.\,\eqref{decoupled}, according to the explicit formula 
\begin{equation}
\label{S}
S = \sum_{n=1}^{2N}\tau_3 \ket{\psi_n}\bra{\psi_n}\tau_3.
\end{equation}  
Together with the $\tau_3$-PH property, the above expression implies that $[G,\tau_3 S]=0$. Hence, by the spectral theorem, there exists a simultaneous eigenbasis of $G$ and $\tau_3 S$ which is orthonormal with respect to the $S$-inner product. The basis $\{\ket{\psi_n}\}_{n=1}^{2N}$ is precisely this basis.  

To better appreciate the role played by of dynamical stability, it is useful to notice that \( \tr S =\sum_{n=1}^{2N} r_n^{-1} \) in terms of the \textit{Krein phase rigidities} $r_n\equiv \braket{\psi_n|\psi_n}^{-1}$  of the eigenvectors of $G$. As we showed in \cite{decon}, there exists an $n$, such that $r_n\to 0$ as the system approaches a dynamical instability. Owing to the positivity of $S$, $\tr S\to \infty$ in the same limit and,  since it follows from Eq.\,\eqref{S} that $\tr S^{-1} = \tr \tau_3 S \tau_3 = \tr S$, we conclude that there exists an eigenvalue of $S$ that tends to zero. In other words, \(S\) becomes ill-defined as a dynamically unstable regime is approached.

\section{Restoring number conservation with duality transformations}
\label{dualitysec}
We are now in a position to construct the desired duality transformation. Given that $S$ is a positive-definite matrix, there is a well-defined, positive-definite square root $R\equiv S^{1/2}$. Since $R$ is unique, it inherits several properties from $S$ itself. Firstly, owing to the positive-definiteness, $R$ is necessarily Hermitian. Secondly,  since $S^{-1}=\tau_3 S\tau_3$, taking the unique positive-definite square root of each side implies that $R^{-1}=\tau_3 R \tau_3$. In agreement with the mathematical framework of \cite{aliexact}, $R^{-1}$ is a Hilbert space isomorphism when viewed as a map from $\mathcal{H}$ to $\mathcal{H}_S$. Finally, rewriting $S$ in the form 
\begin{equation}
S = \sum_{n=1}^{N} \tau_3\, (\ket{\psi_n}\bra{\psi_n} + \ket{\overline{\psi}_n}\bra{\overline{\psi}_n})\,\tau_3
\end{equation}
shows that $S^* = \tau_1 S \tau_1$. Again, taking the unique positive-definite square root of each side yields $R^* = \tau_1 S \tau_1$. 

The above properties allow us to conclude that the map $\hat{\Phi}\mapsto \hat{\Theta}_S\equiv R^{-1}\hat{\Phi}=S^{-1/2}\hat{\Phi}$ is a (linear) unitary, canonical transformation \cite{ripka}.  That is, the operators $\hat{\Theta}_S=[b_1,b_1^\dag,\ldots,b_N,b_N^\dag]^T$ satisfy the bosonic commutation relations $[b_i,b_j^\dag ]=\delta_{ij}1_F$, $[b_i,b_j]=0$. By the Stone-von Neumann theorem \cite{hallQM}, there exists a unitary operator $\widehat{U}$ on Fock space, such that $R^{-1}\hat{\Phi} \equiv \widehat{U}\hat{\Phi}\widehat{U}^\dag$. This unitary operator acts on the many-body Hamiltonian as
\begin{equation}
\widehat{H}^D\equiv \widehat{U}\widehat{H}\widehat{U}^\dag =\frac{1}{2}\hat{\Phi}^\dag (\tau_3 G^D )  \hat{\Phi},\quad G^D  = RGR^{-1}.
\end{equation}
Since the effective BdG Hamiltonian $G^D$ is Hermitian, 
\begin{equation*}
G^D{}^\dag = R^{-1}G^\dag R = R^{-1} S G S^{-1} R = RG R^{-1} = G^D ,
\end{equation*}
it follows that the {\em pairing amplitudes $\Delta^D_{ij}$ of $\widehat{H}^D$ vanish}: the dual QBH \(\widehat{H}^D\equiv \widehat{U}\widehat{H}\widehat{U}^\dag\) satisfies $[\widehat{H}^D,\widehat{N}]=0$. One can also check that, under our duality transformation,  the total number operator for the quasiparticles of \(\widehat{H}\) in Eq.\,\eqref{decoupled},  \( \sum_{n=1}^N \widehat{\psi}_n^\dag \widehat{\psi}_n \), is mapped to \(\widehat{N} \).  

It is instructive to notice the fate of the dual Hamiltonian when pairing vanishes to begin with. In this case, \(G\) already commutes with $\tau_3$. Thus, one can construct an orthonormal basis consisting of simultaneous eigenvectors of $G$ given by $\{\ket{\psi_n^\pm}\}_{n=1}^N$, with $\tau_3 \ket{\psi^\pm_n}=\pm \ket{\psi^\pm_n}$ and $\ket{\psi^\pm_n} = \mathcal{C}\ket{\psi^{\mp}_n}$.  A straightforward calculation of the positive-definite metric $S$ in Eq. \eqref{S} shows that \(S= \mathds{1}_{2N}\). Hence, \(R=\mathds{1}_{2N}\) and the dual effective BdG Hamiltonian is $G^D=RGR^{-1} = G$, implying that \emph{number-conserving QBHs are invariant} under our duality map.

 A few important remarks are in order. First, there is a ``trivial'' sense in which any dynamically stable QBH $\widehat{H}$ is unitarily equivalent to a number-conserving dual: simply take the unitary that maps the normal modes $\widehat{\psi}_n$ to the bosonic operators $a_n$. Of course, to construct such a unitary, one must fully diagonalize $G$. In general, however, our duality does \emph{not} return the fully diagonalized Hamiltonian, as evidenced by the fact that arbitrary number-conserving QBHs are a fixed point of the transformation. Furthermore, despite the definition of $S$ including the complete set of eigenvectors of $G$, a full diagonalization of $G$ is \emph{not always needed} to identify $\widehat{H}^D$ in practice. 
 
Second, it is not \textit{a priori} clear how the locality properties of the original Hamiltonian \(\widehat{H}\) transmute into those of $\widehat{H}^D$. In fact, when the locality properties happen to be preserved, it is straightforward to construct $S$. In this case, $S$ must be site-local. Combining this with the fact that $S$ must be positive-definite and a proper canonical transformation, we have the following Ansatz,
\begin{eqnarray}
\label{localS}
&&S =\sum_{j=1}^N\ket{j}\bra{j}\otimes S_j ,\\
&&S_j \equiv \cosh(\theta_j)\mathds{1}_2+\sinh(\theta_j)\left[\cos(\phi_j)\sigma_x+\sin(\phi_j)\sigma_y\right] .\nonumber
\end{eqnarray}
The condition that $S$ must block-diagonalize $G$ allows one to solve for the parameters $\{\theta_j, \phi_j\}$ in a straightforward way. If no such strictly local transformation exists, one can look at ``quasi-local'' (e.g., two-site-local) transformations and parameterize in an analogous way. Carrying this procedure on allows one to construct $S$ \textit{without} full diagonalization. The examples that follow illustrate that \(\widehat{H}^D\) may or may not be of finite range even if \(\widehat{H}\) is; we will see that such an Ansatz suffices for the bosonic Kitaev chain.

Lastly, it is natural to ask whether an analogous matrix to $S$ may be constructed for quadratic \textit{fermionic} Hamiltonians and, if so, whether it also allows to remove pairing. Two lines of reasoning could be envisioned: (i) since $\tau_3$ is replaced by $\mathds{1}_{2}$ for fermionic SPHs (i.e., $G=H$ is Hermitian, and no instabilities occur), the spectral theorem implies $S=\mathds{1}_{2N}$; (ii) if, on top of Hermiticity, we demand $\tau_3$-PH for a fermionic SPH, then it necessarily commutes with $\tau_3$ and hence cannot contain any pairing to begin with. Either ways, the construction used for bosons does \textit{not} allow for pairing removal in the fermionic case.

\subsection{Example 1: A gapped harmonic chain} Consider the following one-dimensional Hamiltonian under periodic boundary conditions (PBCs),
\begin{equation}
\label{ghcham}
\widehat{H} = \sum_{j=1}^N\Big(\frac{p_j^2}{2m} + \frac{\Co}{2} x_j^2 + \frac{\Cnn}{2}\left( x_{j+1}-x_j\right)^2\Big) ,
\end{equation}
where $x_j$ ($p_j$) is the position (momentum) operator at site $j$, $m>0$ is the uniform mass, and \mbox{$\Co,\Cnn\geq 0$} are onsite and nearest-neighbor (NN) stiffness constants, respectively. Defining $\Omega\equiv \sqrt{(2\Cnn+\Co)/m}$, $J\equiv \Cnn/m\Omega$, and the bosonic annihilation operator $a_j \equiv \sqrt{m\Omega/2}\left(x_j+ip_j/m\Omega\right)$ gives the QBH
\begin{equation}
\label{original1}
\widehat{H} = \!\sum_{j=1}^N \Omega\Big( a_j^\dag a_j + \frac{1}{2}\Big) 
- \frac{J}{2}\Big(a_{j+1}^\dag a_j + a_{j+1}^\dag a_j^\dag + \text{H.c.}\Big),
\end{equation}
which explicitly breaks $a$-boson number conservation. We focus in the following on the dynamically stable parameter
regime\footnote{When $C_\text{o}=0$, $\widehat{H}$ is the standard one-dimensional phonon chain, that possesses a free-particle excitation at zero energy corresponding to the conserved total momentum. This manifests as a loss of diagonalizability in $G$ and hence the onset of instability. } \(\Co>0, \Cnn\geq 0\).

By moving to the Fourier basis, that is, $b_k\equiv N^{-1/2}\sum_{j=1}^N e^{-ikj}a_j$, with $k$ in the first Brillouin zone (BZ), one obtains the Bloch effective BdG Hamiltonian \(G_k\), which is a \(k\)-dependent \(2\times 2\) matrix. The bosonic normal modes \(\beta_k, \beta_k^\dagger\) are calculated by the method sketched in the background section, leading to 
\begin{eqnarray*}
\ \widehat{H}=\!\!\sum_{k\in\text{B.Z.}} \omega_k \beta_k^\dag \beta_k,
\ \ \omega_k\equiv[(\Co+4\Cnn\sin^2(k/2))/m]^{1/2}.
\end{eqnarray*}
Recall that our duality is designed to remove the pairing terms in the effective BdG Hamiltonian. For the $2\times 2$ matrix $G_k$, 
this goal is equivalent to diagonalizing $G_k$. Accordingly, the induced many-body transformation is simply \(\beta_k\mapsto b_k\), that is, $\widehat{H}^D = \sum_{k\in\text{B.Z.}}\omega_k b_k^\dag b_k$,
which indeed commutes with \(\widehat{N}\). In real space, 
\begin{eqnarray*}
\widehat{H}^D = \sum_{j=1}^N\sum_{r=1}^{N-j} \left( K^D_{r}a_{j+r}^\dag a_{j} + \text{H.c.}\right)+K_0^D\widehat{N},
\end{eqnarray*}
in terms of \(K^D_r \equiv N^{-1}\sum_{k\in\text{B.Z.}} \omega_k e^{ikr}\). Hence, the effect of pairing in the original NN Hamiltonian of Eq.\,\eqref{original1} is mimicked in $\widehat{H}^D$ by rapidly decaying (see below and Fig.\,\ref{Kvr}) but non-finite-range hopping amplitudes.

Notice that $K^D_r=\Omega \delta_{r0}$ in the number-conserving limit $\Cnn=m\Omega J=0$ [see Eq.\,\eqref{original1}], and so the original and the dual system coincide, as expected on general grounds. By contrast, as soon as $\Cnn>0$ and $\Co>0$, the hopping amplitudes of the dual model are no longer of finite range. The exact value of the dual hopping amplitudes in the limit $N\to\infty$ can be evaluated analytically at the point $\Co=0$, where the gap closes and the system is no longer dynamically stable. By letting $\Omega_\text{nn}\equiv 2\sqrt{\Cnn/m}$, we find 
\begin{equation}
\label{exact}
K^\text{TL,0}_r \equiv \frac{1}{2\pi}\int_{-\pi}^\pi \left(\omega_k|_{\Co=0}\right) e^{ikr}\,dk = \frac{2}{\pi}\frac{\Omega_\text{nn}}{1-4r^2} .
\end{equation} 
As seen in Fig.\,\ref{Kvr}(a), this $1/r^2$-type limiting value appears to bound the exact hopping amplitudes (calculated numerically) for finite system size. So, how well does $\widehat{H}^D$ approximates $\widehat{H}$ if we truncate the hopping range  to some finite value $r_\text{max}\equiv \varrho$? Let $\widehat{H}^D(\varrho)$ denote the truncated Hamiltonian and $\omega^\varrho_k$ the corresponding band structure. A plot of $\omega^\varrho_k$ is given for $\varrho=0,1,2,3$ in Fig.\,\ref{trunc}. Note that the quasiparticle gap is present for each $\varrho$ despite the lack of pairing-like terms in the truncated Hamiltonian.

\begin{figure}[t]
\includegraphics[width=0.9\columnwidth]{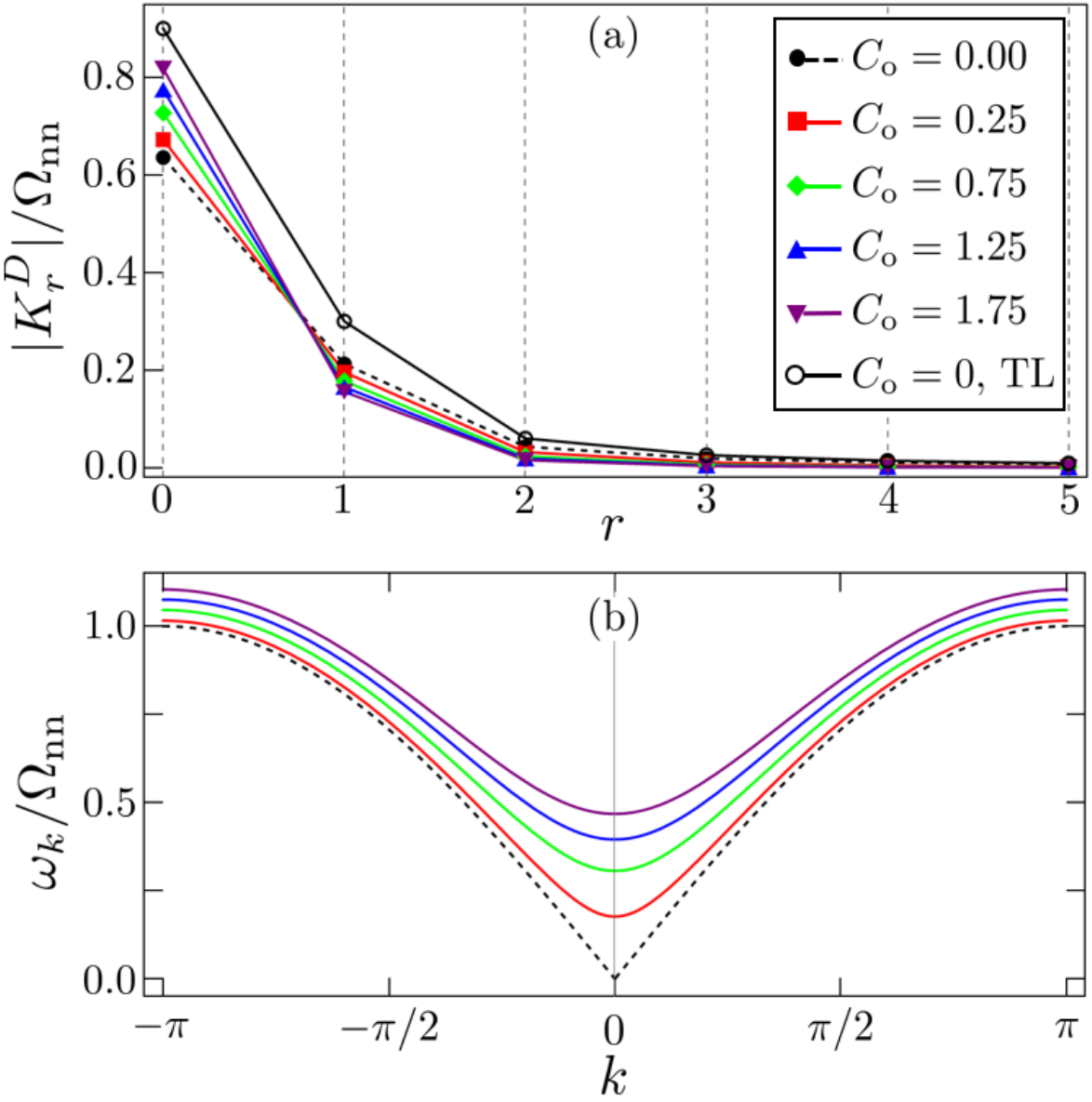}
\vspace*{-4mm}
\caption{(Color online) 
(a) Rescaled hopping strength $|K^D_r|/\Omega_\text{nn}$, for varying onsite stiffness $\Co$. In all cases $m=1,\Cnn=2$ and $N=30$. The exact expression for the hopping amplitude in the thermodynamic limit [TL, Eq. (\ref{exact})] is also shown for $\Co=0$. (b) The band structure $\omega_k$, with the same normalization and same parameter values, for varying $\Co$. The color coding is the same as in (a).  Note that the duality transformation is not strictly valid for $\Co=0$ due to loss of diagonalizability.}
\label{Kvr}
\end{figure}

\begin{figure}[t]
\begin{center}
\includegraphics[width=0.87\columnwidth]{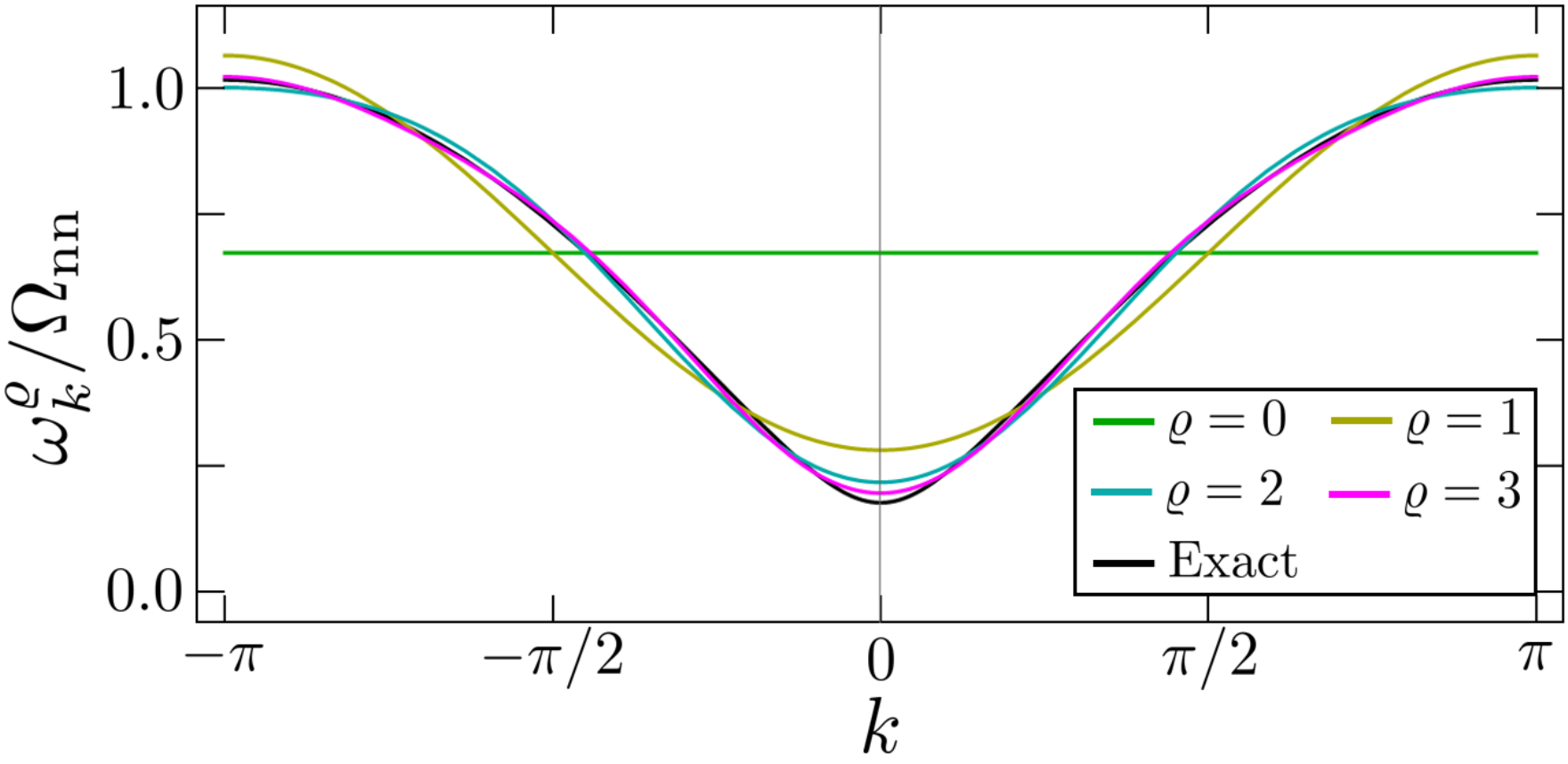}
\end{center}
\vspace*{-4mm}
\caption{(Color online) 
The band structure of the Hamiltonian $\widehat{H}^D(\varrho)$ with the real-space coupling range truncated at $\varrho=0,1,2,3$. The exact band structure is also shown for comparison. All remaining parameters are the same as in Fig. \ref{Kvr}.}
\label{trunc}
\end{figure}

We emphasize that the procedure we carried out in this example generalizes to \emph{any} translationally-invariant,  dynamically stable QBH without internal degrees of freedom, regardless of the space dimensionality. Given that the band structure is delocalized in momentum space, the hopping amplitudes of $\widehat{H}^D$ will generically be ``short-range'', i.e., exponentially decaying in real space. Truncation will then produce a finite-range number-conserving QBH which is approximately isospectral to the original Hamiltonian with small error.

\subsection{Example 2:  A bosonic analogue of Kitaev's Majorana chain}
In this example our duality transformation does not change the range of the hopping amplitudes. The original model is a bosonic chain motivated by a certain analogy to the fermionic Majorana chain of Kitaev \cite{clerkBKC,decon}. The QBH is of the form $\widehat{H}\equiv \widehat{H}_O + s\widehat{W}(\varphi)$, with
\begin{eqnarray}
\label{BKC}
\begin{split}
\widehat{H}_O &\equiv \!\!& \frac{1}{2}\sum_{j=1}^{N-1} \left(it a_{j+1}^\dag a_j+i\Delta a_{j+1}^\dag a_{j}^\dag + \text{H.c.}\right),
\cr
\widehat{W}(\varphi) &\equiv \!\!\!& \frac{1}{2}\left(it e^{i\varphi} a_{1}^\dag a_N+i\Delta e^{i\varphi} a_{1}^\dag a_{N}^\dag + \text{H.c.}\right).
\end{split}
\end{eqnarray}
Here, $\widehat{H}_O$ represents the system under open boundary conditions (OBCs), $\widehat{W}(\varphi)$ introduces $\varphi$-twisted boundary conditions ($\varphi$-TBCs), and the parameters $t,\Delta>0$, $s\in[0,1]$, $\varphi\in[0,\pi]$. Analytical solutions reveal that in the hopping-dominated regime, $t>\Delta$, the system is \emph{dynamically stable} for both OBCs and $\pi/2$-TBCs with $s=1$;   likewise, numerics indicate stability in a small region in the $(\varphi,s)$-plane surrounding the line $\varphi=\pi/2$ for all $N$ in addition to the line $s=0$ for $N$ even, which shrinks exponentially as $N$ increases \cite{decon}. Instead, the system is dynamically unstable in the pairing-dominated regime,  $\Delta>t$  (see Fig.\,\ref{dynstab} for a representative dynamical phase diagram as a function of boundary parameters). Since our duality transformation is only defined for dynamically stable QBHs, we limit ourselves to the case $t>\Delta$. 

\begin{figure}[t]
\begin{center}
\includegraphics[width=.35\textwidth]{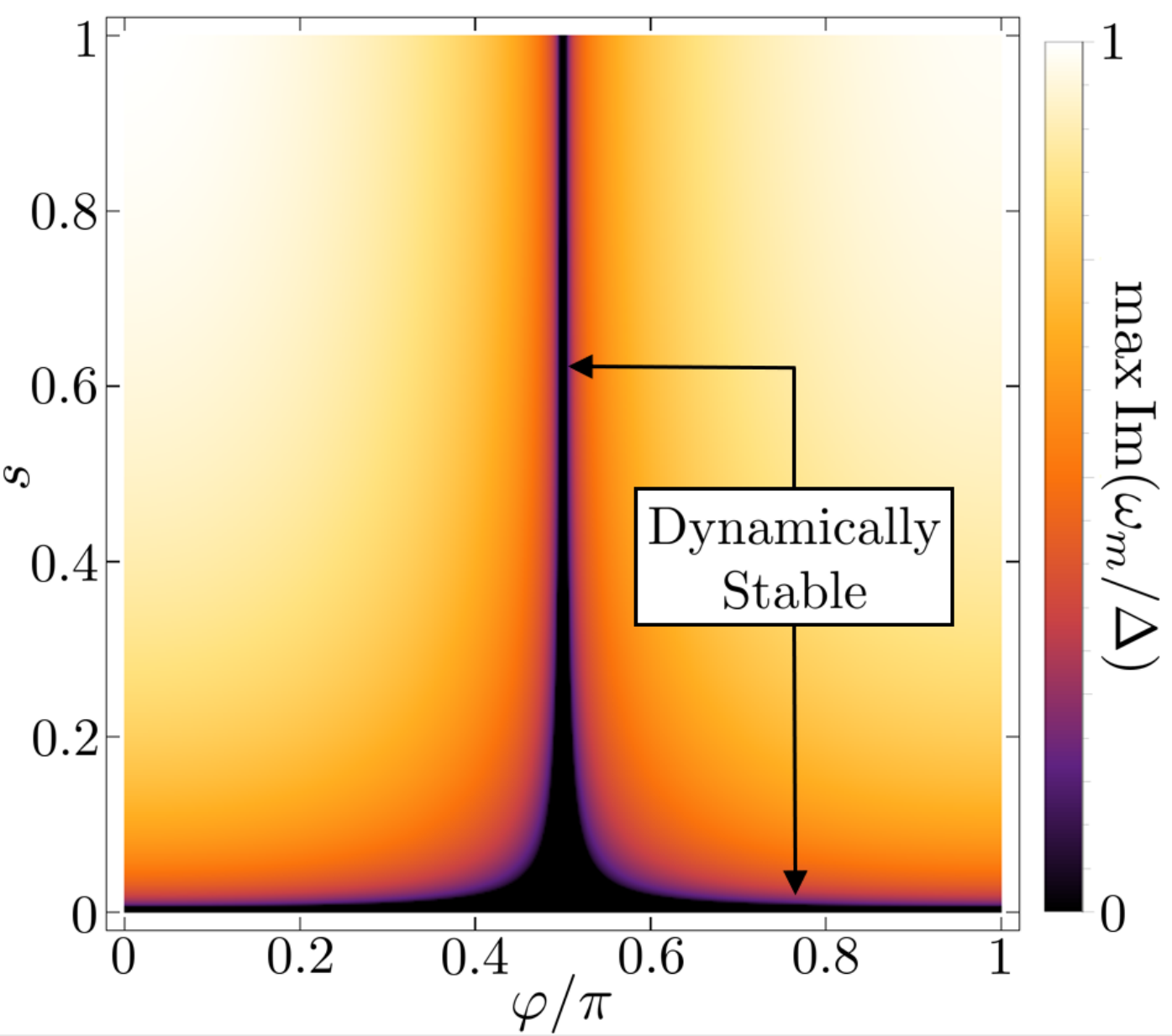}
\end{center}
\vspace*{-4mm}
\caption{(Color online) 
A dynamical phase diagram for the Hamiltonian in Eq. \eqref{BKC} with $t=1, \Delta=0.25$, $N=20$. Here $\omega_m$ denotes the $m$-th eigenvalue of the effective SPH. Note that the regions of stability ($\max\textup{Im}(\omega_m)=0$) are concentrated around the lines $\varphi=\pi/2$ and $s=0$.}
\label{dynstab}
\end{figure}

Utilizing the analytical solutions for the basis $\{\ket{\psi_n}\}$ derived in \cite{decon}, we can determine the positive-definite matrix $S$ for $s=1$ and $\varphi=\pi/2$, by using Eq.\,\eqref{S}. We find
\begin{eqnarray*}
S_{\pi/2}(r) &\!\!\!=\!\!\!& \sum_{j=1}^N\ket{j}\bra{j}\otimes S_j(r) ,
\cr
S_j(r) &\!\!\!\equiv\!\!\!& \begin{bmatrix}
\cosh[2(j-j_0)r] & -\sinh[2(j-j_0)r] 
\\
-\sinh[2(j-j_0)r]  & \cosh[2(j-j_0)r] 
\end{bmatrix} ,
\end{eqnarray*}
with $j_0\equiv (N+2)/2)$ and $r\equiv (1/2)\ln\left[(t+\Delta)/(t-\Delta)\right]$. Noting that $R^{-1}(r)\equiv S^{-1/2}_{\pi/2}(r) = S_{\pi/2}(-r/2)$, it follows that the desired duality transformation is given by 
\begin{equation}
a_j\mapsto \cosh[(j-j_0)r]a_j + \sinh[(j-j_0)r]a_j^\dag,
\label{BKmap}
\end{equation}
which yields the number-conserving dual QBH 
\begin{equation}
\widehat{H}^D = \frac{i\tilde{t}}{2}\sum_{j=1}^{N-1}\! \left(a_{j+1}^\dag a_j - \text{H.c.}\right) -\frac{s\tilde{t}}{2}\left(a_1^\dag a_N + \text{H.c.}\right),
\label{BKCD}
\end{equation}
where $\tilde{t}\equiv\sqrt{t^2-\Delta^2}$. The first term in Eq. (\ref{BKCD}), which we shall denote $\widehat{H}^D_O$, is the image, under the transformation, of the bulk Hamiltonian $\widehat{H}_O$, whereas the second term is the image of the boundary term $s\widehat{W}(\pi/2)$. We see that as $\Delta\to t$, the QBH $\widehat{H}^D$  approaches the zero Hamiltonian and \(\widehat{H}\) approaches a dynamical instability. We also see that, although constructed for $s=1$, our duality holds unchanged for \emph{any} $s\in[0,1]$ at $\varphi=\pi/2$. Thus, by leveraging exact solutions at a single point, we have constructed the duality map for a non-trivial region in parameter-space.

The special case of this duality  with \(s=0\) was discovered by invoking an analogous Ansatz to that in Eq.\,\eqref{localS} in \cite{clerkBKC}, upon leveraging insight about the physics of squeezing. In particular, full diagonalization of $G_O$ was not necessary. As it turns out, for OBCs only, our duality maps $\widehat{H}_O$ to $\widehat{H}^D_O$, \emph{regardless} of the choice of $j_0$ in Eq. (\ref{BKmap}); $j_0$ may even vary spatially as a function of $j$. This parametric freedom, also noticed to an extent in \cite{clerkBKC}, can be explained within our framework as a consequence of the chiral ($+/-$) symmetry in the quasi-particle \textit{excitation} energy spectrum of $\widehat{H}_O$ \cite{decon}. Moreover, while we have not taken this route, this freedom allows for another method to construct $S_{\pi/2}(r)$ without a full diagonalization, by solving for $j_0$ under the constraint that $\widehat{W}(\pi/2)$ must be mapped to a number-conserving boundary condition. Here again the Ansatz in Eq.\,\eqref{localS} greatly simplifies the computation.

Notably, a proposal to use $\widehat{H}_O$ for generating multipartite entangled states was also put forward in \cite{clerkBKC}, taking advantage of the fact that $\widehat{H}^D_O$ can be thought of as a beam-splitter network and, thanks to the locality properties of the mapping $\widehat{H}_O \leftrightarrow \widehat{H}^D_O$, non-trivial entanglement properties of output states remain unchanged in the process. Through the lens of our more general duality transformation, such an application is always possible when the map is locality-preserving as in Eq. (\ref{localS}) -- in which case, one may show that the duality takes precisely the form of a generalized local squeezing transformation. As for more general quasi-local dualities, for which the dual couplings affect, exactly or approximately, a finite number of subsystems, one can still consider generating states that are non-trivially entangled relative to a suitably generalized notion of entanglement. Specifically, by considering entanglement relative to a coarse-grained (e.g., bi-local) lattice partition to accommodate the locality structure of the duality transformation \cite{GE}, one can imagine generating generalized entangled states with easily implementable dynamically stable bosonic systems.

\section{Duality and topological invariants} QBHs can display non-trivial bands characterized in terms of topological invariants. On the one hand, the topological invariants of number-conserving QBHs coincide with the well-established ones for fermions\cite{ProdanBook}, even though the many-body interpretation of these quantities can change drastically \cite{Xu}. On the other hand, for number-non-conserving QBHs the appropriate topological invariants are defined with respect to the indefinite metric of the Krein space ${\mathscr{K}}_{\tau_3}$ on which the effective BdG Hamiltonian acts \cite{Shindou}. Here, we will show how our duality handles the translation between these, \textit{a priori} very different, bulk invariants. In particular, this result makes it possible to use the bulk-boundary correspondence for number-non-conserving bosons \cite{baldespeano}, along with the standard one for the number-conserving dual, to relate aspects of the respective edge-mode physics, without the need to explicitly compute the duality for OBCs. We also note that a related mapping from \textit{thermodynamically stable} bosonic BdG SPHs to Hermitian ones was identified in \cite{LeinSato}, where it was argued that topological classification of the Hermitian Hamiltonian is equivalent to that of the corresponding pseudo-Hermitian one. With a similar philosophy in mind, we examine the extent to which this holds.

Let us first recall the formula for the indefinite \(\tau_3\)-inner product equivalent of the Berry connection \cite{Shindou}, by focussing on the Abelian case for simplicity. Let $G(\mathbf{k})$ denote an effective BdG Hamiltonian that depends on a vector of parameters $\mathbf{k}$. Suppose further that $G(\mathbf{k})$ is dynamically stable, with a complete basis of eigenstates $\ket{n(\mathbf{k})}$ satisfying $\braket{n(\mathbf{k})|\tau_3|m(\mathbf{k})}=\lambda_n \delta_{nm}$, $\lambda_n$ being either $+1$ or $-1$. Following the usual assumptions of adiabatic evolution for the dynamics genereted by $G(\mathbf{k})$ \cite{Xu}, one finds that 
\begin{eqnarray}
\label{KB}
A_{\mathscr{K}}(\mathbf{k})\equiv i\lambda_n\braket{n(\mathbf{k})|\tau_3\nabla_\mathbf{k}|n(\mathbf{k})}
\end{eqnarray}
is the connection for parallel transport in  ${\mathscr K}_{\tau_3}$. The dual (Hermitian) BdG Hamiltonian is $G^D (\mathbf{k})\equiv R(\mathbf{k})G(\mathbf{k})R^{-1}(\mathbf{k}),$ where $R(\mathbf{k})$ is the unique positive-definite square root of the metric $S(\mathbf{k})$ defined in Eq.\,\eqref{S}. A basis of eigenvectors of $G^D (\mathbf{k})$ is then given by $\ket{n^D(\mathbf{k})} = R(\mathbf{k})\ket{n(\mathbf{k})}$ and can be chosen to satisfy $\braket{n^D(\mathbf{k})|m^D (\mathbf{k})}=\delta_{nm}$. Thus, the  usual Berry connection reads
\begin{equation*}
A_{\mathscr{B}}(\mathbf{k}) =  i\braket{n^D (\mathbf{k})|\nabla_\mathbf{k}|n^D (\mathbf{k})}.
\end{equation*}

What is the relationship between these two connections and, more importantly, the associated topological invariants? Utilizing the explicit form of the eigenstates of $G^D (\mathbf{k})$, the Berry connection can be written as
\begin{eqnarray*}
A_{\mathscr{B}}(\mathbf{k}) &\!\!\!=\!\!\!& i\braket{n(\mathbf{k})|R(\mathbf{k})\left(\nabla_\mathbf{k} R(\mathbf{k})\right)|n(\mathbf{k})}
\\ & \!\!\!+\!\!\! & i \braket{n(\mathbf{k})|S(\mathbf{k})\nabla_\mathbf{k}|n(\mathbf{k})}.
\end{eqnarray*}
Since we can ensure that $\ket{n(\mathbf{k})}$ is a simultaneous eigenvector of both $G$ and $\tau_3 S(\mathbf{k})$, with $\tau_3 S(\mathbf{k})\ket{n(\mathbf{k})} = \lambda_n \ket{n(\mathbf{k})}$, the second term in the above expression can be rewritten as $i \braket{n(\mathbf{k})|S(\mathbf{k})\nabla_\mathbf{k}|n(\mathbf{k})} = i \lambda_n\braket{n(\mathbf{k})|\tau_3\nabla_\mathbf{k}|n(\mathbf{k})} = A_\mathscr{K}(\mathbf{k}).$ In conclusion,
$$ A_\mathscr{B}(\mathbf{k})-A_\mathscr{K}(\mathbf{k})= i\braket{n(\mathbf{k})|R(k)\left(\nabla_\mathbf{k} R(\mathbf{k})\right)|n(\mathbf{k})}. $$
One can determine from this identity how various topological invariants are related. One can show (see Appendix) that \(A_\mathscr{B}=A_\mathscr{K}\) if $[R(\mathbf{k}),\nabla_\mathbf{k} R(\mathbf{k})]=0$. Despite lacking (as of now) a clear physical interpretation, this condition gives a straightforward way for determining when these connections, and hence the associated invariants, coincide. 

\section{Duality and quantum simulation}
\label{apps}
Analog simulators aim to implement a target ``surrogate'' 
Hamiltonian in an analog (as opposed to gate-based) fashion \cite{nori}.
Suppose we wish to realize a set of $N$, parametrically driven (or paired) bosons, described by a target Hamiltonian of the form in Eq. (\ref{genham}), with $\Delta_{ij}$ not all zero. While there exists ways to implement these terms in experimental settings \cite{clerkBKC,hafazi}, the need for precisely tuned parametric amplification introduces extra complications. As we have seen, in a dynamically stable regime, our duality transformation can unitarily map the original Hamiltonian to one that lacks any driving terms. When this transformation is sufficiently local (for instance, as in Eq. \eqref{localS}), one can experimentally access properties of the original system by directly implementing the dual. If $\widehat{H}^D$ possesses arbitrary-range couplings that drop off exponentially with distance, finite-range truncation can faithfully reproduce the spectral properties of the original system. Thus, generically, the stable dynamics of parametrically driven systems can be faithfully realized in a system that comprises \emph{only} suitably adjusted short-range hopping amplitudes. 

This further leads naturally to the possibility of identifying Hermitian, number-conserving QBHs whose spectral properties well approximate (or even exactly replicate) those of truly {\em non-Hermitian, $\mathcal{P}\mathcal{T}$-symmetric} systems, as actively investigated across  photonic, optomechanical, and cavity QED settings \cite{PT,wiersig}. Motivated by the fact that typical implementations in open (dissipative) systems with balanced gain and loss also entail unavoidable introduction of noise, a related question was addressed in \cite{clerkPT}: under which conditions can a target $\mathcal{P}\mathcal{T}$-symmetric Hamiltonian be realized in a non-dissipative quantum system of free bosons? The authors found a class of $\mathcal{P}\mathcal{T}$-symmetric systems whose dynamics can be unitarily mapped to those of a non-dissipative QBH. Our duality transformation allows us to say something more about this class: since the $\mathcal{P}\mathcal{T}$-unbroken phases of these models must possess an entirely real spectra \cite{decon}, the resulting QBHs will be dynamically stable. Hence, the $\mathcal{P}\mathcal{T}$-unbroken regimes of the target non-Hermitian system can be faithfully recreated \emph{without} the need for dissipation or parametric driving in the bosonic dual number-conserving system. This not only greatly reduces the complexity of experimental implementation but, by removing amplification mechanisms, it prevents the extreme sensitivity to imperfections that is distinctive of these systems.

\newpage

\section{Conclusion}
\label{conclusion}
We have shown terms that break number conservation in dynamically stable QBHs can always be removed by a Hamiltonian-dependent, but fully-specified, duality transformation. Conceptually, our analysis fully exposes the significance that dynamical stability carries for non-interacting bosonic systems, further highlighting key differences from their fermionic counterparts. Identifying the most general mathematical conditions under which duality transformations to number-conserving systems may be constructed without requiring complete diagonalization, or may be guaranteed to obey specified quasi-locality constraints, are well-worth related questions for further investigation. From a practical perspective, our dualities may find immediate application in analog quantum simulation, by providing new means for robustly realizing $\mathcal{P}\mathcal{T}$-symmetric systems and their rich physics in number-conserving Hamiltonians with only hopping terms. Since, once calculated, our dualities can be repurposed by applying them to the same original QBH modified by interactions and disorder or, possibly, couplings to an external (e.g., Markovian) environment, they may ultimately prove a valuable tool for pushing the simulation in yet unexplored dynamical regimes.

\acknowledgments
Work at Dartmouth was partially supported by the US NSF through Grant No. PHY-1620541, the US DOE, Office of Science, Office of Advanced Scientific Computing Research, Accelerated Research for Quantum Computing program, and the Constance and Walter Burke Special Projects Fund in QIS. E.C. acknowledges partial support from a 2019 seed grant from SUNY Poly Research Office.

\section{Appendix: Sufficient condition for the equality of Berry connections} We derive here the condition for the Berry phase associated to a dynamically stable, number-non-conserving QBH with effective SPH $G$ to coincide with the standard Berry phase associated with the Hermitian dual $G^D$. As shown in the main text, 
\begin{equation*}
A_\mathscr{B}(\mathbf{k})-A_\mathscr{K}(\mathbf{k})=i\braket{n(\mathbf{k})|R(k)\left(\nabla_\mathbf{k} R(\mathbf{k})\right)|n(\mathbf{k})} .
\end{equation*}
Now, if we assume that $[R(\mathbf{k}),\nabla_\mathbf{k} R(\mathbf{k})]=0$, then $R(k)\left(\nabla_\mathbf{k} R(\mathbf{k})\right) = \frac{1}{2}\nabla_\mathbf{k} S(\mathbf{k})$. Furthermore,
\begin{eqnarray*}
&&\braket{n(\mathbf{k})|\left(\nabla_\mathbf{k} S(\mathbf{k})\right)|n(\mathbf{k})} = \nabla_\mathbf{k}\left(\braket{n(\mathbf{k})|S(\mathbf{k})|n(\mathbf{k})}\right) - \\
&& \left(\nabla_\mathbf{k} \bra{n(\mathbf{k})}\right)S(\mathbf{k})\ket{n(\mathbf{k})} - \bra{n(\mathbf{k})}S(\mathbf{k}) \nabla_\mathbf{k}\ket{n(\mathbf{k})} .
\end{eqnarray*}
The first term is zero by virtue of $\ket{n(\mathbf{k})}$ providing an $S(\mathbf{k})$-orthonormal basis. Again, utilizing the fact that $\tau_3 S(\mathbf{k})\ket{n(\mathbf{k})} = \lambda_n \ket{n(\mathbf{k})}$, we have
\begin{eqnarray*}
&&\braket{n(\mathbf{k})|\left(\nabla_\mathbf{k} S(\mathbf{k})\right)|n(\mathbf{k})} =  - \lambda_n\left(\nabla_\mathbf{k} \bra{n(\mathbf{k})}\right)\tau_3\ket{n(\mathbf{k})} - \\
&& \lambda_n\bra{n(\mathbf{k})}\tau_3\nabla_\mathbf{k}\ket{n(\mathbf{k})} = -\lambda_n\nabla_\mathbf{k}\left(\braket{n(\mathbf{k})|\tau_3|n(\mathbf{k})}\right) = 0 ,
\end{eqnarray*}
where we have used $\braket{n(\mathbf{k})|\tau_3|n(\mathbf{k})}=\lambda_n$. Altogether,
\begin{eqnarray*}
A_\mathscr{B}(\mathbf{k})-A_\mathscr{K}(\mathbf{k}) &\!\!\!=\!\!\!&i\braket{n(\mathbf{k})|R(k)\left(\nabla_\mathbf{k} R(\mathbf{k})\right)|n(\mathbf{k})} \\
&\!\!\!=\!\!\!& \frac{i}{2}\braket{n(\mathbf{k})|\left(\nabla_\mathbf{k} S(\mathbf{k})\right)|n(\mathbf{k})} = 0.
\end{eqnarray*}
We conclude that if $[R(\mathbf{k}),\nabla_\mathbf{k} R(\mathbf{k})]=0$, then $A_\mathscr{B}(\mathbf{k})=A_\mathscr{K}(\mathbf{k})$, as stated.\hfill$\Box$

\end{document}